\documentclass[twocolumn]{aastex61}
\usepackage{amsmath,amstext}
\usepackage[T1]{fontenc}
\usepackage{apjfonts} 
\usepackage[figure,figure*]{hypcap}

\shorttitle{Untwisting Jets}
\shortauthors{Jiajia Liu et al.}

\begin{document}

\title{Untwisting Jets Related to Magnetic Flux Cancellation}
\author{Jiajia Liu}
\email{jj.liu@sheffield.ac.uk}
\affiliation{Solar Physics and Space Plasma Research Center (SP2RC), School of Mathematics and Statistics, The University of Sheffield, Sheffield S3 7RH, UK}

\author{Robert Erd{\'e}lyi}
\affiliation{Solar Physics and Space Plasma Research Center (SP2RC), School of Mathematics and Statistics, The University of Sheffield, Sheffield S3 7RH, UK}
\affiliation{Department of Astronomy, E\"{o}tv\"{o}s Lor\'{a}nd University, Budapest, P\'{a}zm\'{a}ny P. s\'{e}t\'{a}ny 1/A, H-1117, Hungary}

\author{Yuming Wang}
\affiliation{CAS Key Laboratory of Geospace Environment, Department of Geophysics and Planetary Sciences, University of Science and Technology of China, Hefei, Anhui 230026, China}

\author{Rui Liu}
\affiliation{CAS Key Laboratory of Geospace Environment, Department of Geophysics and Planetary Sciences, University of Science and Technology of China, Hefei, Anhui 230026, China}

\begin{abstract}
The rotational motion of solar jets is believed to be a signature of the untwisting process resulting from magnetic reconnection, which takes place between twisted closed magnetic loops (i.e., magnetic flux ropes) and open magnetic field lines. The identification of the pre-existing flux rope, and the relationship between the twist contained in the rope and the number of turns the jet experiences, are then vital in understanding the jet-triggering mechanism. In this paper, we will perform a detailed analysis of imaging, spectral and magnetic field observations of four homologous jets, among which the fourth one releases a twist angle of 2.6$\pi$. Non-linear force free field extrapolation of the photospheric vector magnetic field before the jet eruption presents a magnetic configuration with a null point between twisted and open fields - a configuration highly in favor of the eruption of solar jets. The fact that the jet rotates in the opposite sense of handness to the twist contained in the pre-eruption photospheric magnetic field, confirms the unwinding of the twist by the jet's rotational motion.  Temporal relationship between jets' occurrence and the total negative flux at their source region, together with the enhanced magnetic submergence term of the photospheric Poynting flux, shows that these jets are highly associated with local magnetic flux cancellation.
\end{abstract}


\section{Introduction} \label{intro}
Helical structures and motions occur frequently in the solar atmosphere. They could be observed in erupting filaments \citep[e.g.][]{RustD2005, Alexander2006, LiuR2007, Gilbert2007}, sigmoids \citep[e.g.][]{Titov1999, Green2007, ZhangJ2012}, tornadoes \citep[e.g.][]{LiuJ2012, Wedemeyer-Bohm2012, WangW2017}, flares \citep[e.g.,][]{Titov1999}, coronal mass ejections \citep[CMEs,][]{Dere1999, Low2001, Chen2011} and even magnetic clouds in the interplanetary space \citep[e.g.,][]{WangY2015}. As a most likely representation of twisted magnetic field, they are widely believed to play important roles in storing free magnetic energy, resulting in torsional waves or instabilities, and further transferring magnetic energy into thermal/kinetic energies \citep[e.g.,][]{Aschwanden2004, Jibben2004, Jess2009, Falconer2006}

Large-scale solar jets \citep[surges or macro-spicules,][]{Roy1973, Bennett2015, Gyenge2015, Kiss2017}, with different dominant temperature and different manifestation in different wavelength, can either be helical or straight \citep[e.g.,][]{Pariat2015}. Early imaging and spectral observations have demonstrated that rotational motion can be found in many H$\alpha$ surges and X-ray jets \citep[e.g.][]{Canfield1996, Alexander1999}. Recent imaging observations using state-of-art facilities with high temporal and spatial resolution have shown more detailed information about the helical structure of UV/EUV/X-ray jets \citep[e.g.][]{LiuW2009, Cheung2015, LiuJ2014, LiuJ2015, LiuJ2016}. In the case of off-limb jets with significant rotational motions, using only imaging observations might help us investigate some of their properties including axial and rotational speeds. However, it is usually hard to study the direction of their rotational motion due to the complication caused by the line-of-sight (LOS) integration effect \citep[e.g.,][]{LiuJ2014}. In the case of on-disk jets, it is almost impossible for imaging observations to investigate their rotational motions \citep[e.g.,][]{LiuJ2016a}. In such cases, spectral observations can assist and give clues of rotational motion from different Doppler velocities at different parts of jets \citep[e.g.,][]{Scullion2009, Curdt2011}. However, these kind of observations on solar jets are still scarce.

Theory interprets the rotational motion of jets as a result of the untwisting process after magnetic reconnection \citep[for reviews see, e.g.,][]{Shibata1996, Raouafi2016}. A newly emerging or pre-existing closed flux system reconnects with the ambient open magnetic field, during which twists contained in the closed flux system could be passed into the open field. This scenario of the untwisting process in solar jets has been suggested and confirmed by a number of MHD simulations \citep[e.g.,][]{Shibata1986, Moreno-Insertis2008, Pariat2009, Fang2014, Lee2014, Cheung2015}, some of which also show the close relationship between the kink instability of the twisted fields and the initiation of the magnetic reconnection \citep[e.g.,][]{Moreno-Insertis2008, Pariat2009}. Direct simultaneous observation of solar jets and the underneath magnetic field will enable us to perform the comparison between the number of turns a jet rotates and the twist contained in the pre-eruption magnetic flux rope, allowing us to further examine the above theory and the relationship between the eruption of solar jets and kink instabilities.

Flux emergence \citep[e.g.,][]{Murray2009, Fang2014}, cancellation \citep[e.g.,][]{Roussev2001} and rotational/shearing motion \citep[e.g.,][]{Pariat2009, YangL2013} at the footpoint region are the most common photospheric processes related to the eruption of jets either in simulations. Thanks to the increasing number of simultaneous observations of jets and their footpoint regions, the answer of the question whether all the mechanisms above are possible in the real solar atmosphere, is now much clearer \citep[e.g.,][]{Brooks2007, Chifor2008, Guo2013, Innes2016, Nelson2016}. Recent works show a wealth of observational evidences of (recurrent) twisting jets introduced by moving/shearing magnetic features \citep[e.g.][]{Schmieder2013, ChenJ2015}. In our latest work \citep{LiuJ2016}, we performed a detailed analysis of the magnetic and energetic characteristics of recurrent homologous jets. Combined study on the occurrence of jets and the evolution of corresponding photospheric magnetic field, has shown how the emerging process of the magnetic field introduces free magnetic energy and affects the eruption of the recurrent jets. No matter which mechanism of the above dominates, it is the combination of the evolution of the magnetic field and energy flow which plays important roles in determine the eruption of solar jets. However, most previous works have been only focused on the magnetic field part of the whole picture. The energy flow at jets' source regions which represents how the magnetic energy is accumulated/dissipated and where it goes or comes from, has been rarely studied.

In this paper, we will conduct a detailed analysis of four homologous recurrent jets to address the above issues. The paper is organized as following: Analysis of simultaneous FUV and EUV imaging observations from {\it SDO}/AIA and {\it IRIS}/SJI are presented in Sect.~\ref{imaging}. Spectral observations from the {\it IRIS} Spectragraph (SG) are employed to study the rotational behavior of the observed jets in Sect.~\ref{spectral}. In Sect.~\ref{magnetic}, we conduct the study of the twist released by the jet, and the twist the underneath magnetic field contains. The relationship between the occurrence of jets and the photospheric magnetic field variation and Poynting flux is discussed in detail in Sect.~\ref{cancel}. We summarize in Sect.~\ref{summary}.

\begin{figure*}[tbh]
\centering
\includegraphics[width=\hsize]{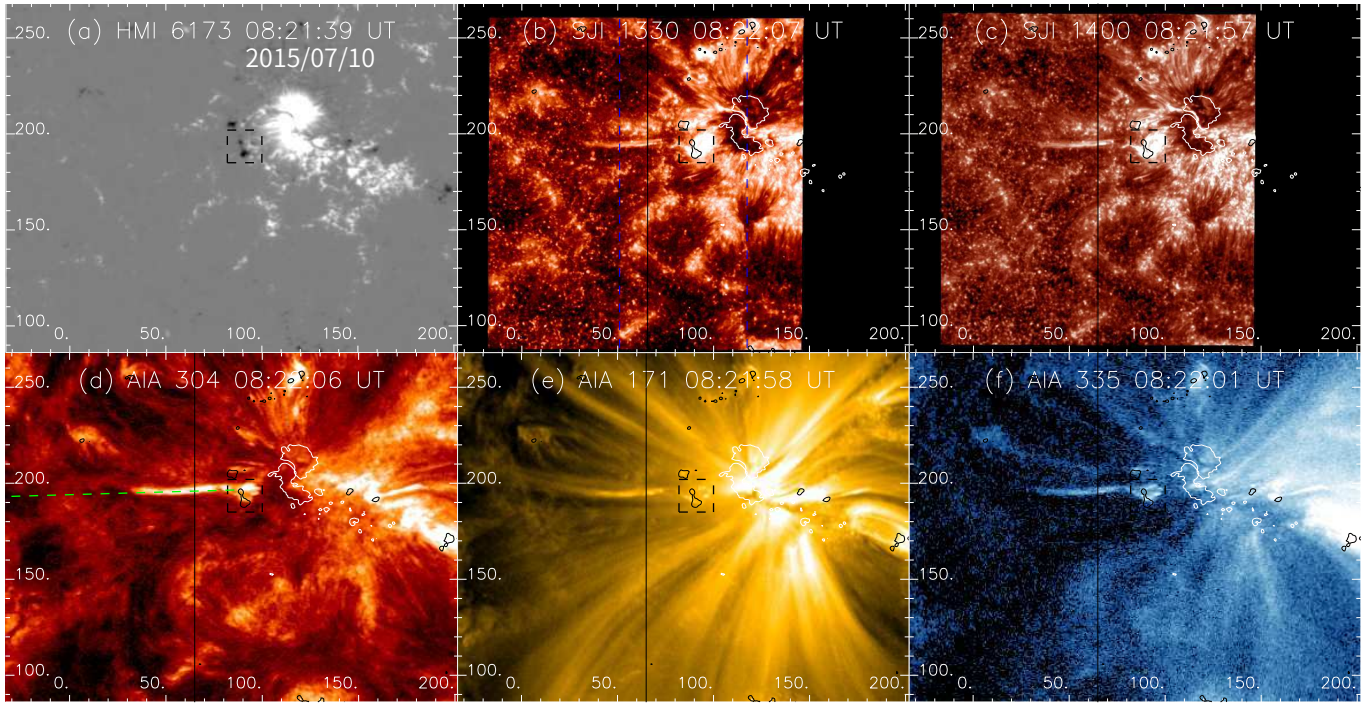}
\caption{{\it SDO}/HMI (panel a), {\it IRIS}/SJI FUV (panel b and c) and {\it SDO}/AIA EUV observation (panel d to f) of the jet at around 08:22 UT July 10th 2015. Dashed boxes surround the source region of the jet with negative polarity, which is also shown as black solid contours with a LOS magnetic field level of -150 G in panel (b) to (f). White contours in panel (b) to (f) depict a LOS magnetic field level of 800 G. Black vertical line in these panels stand for the position of the {\it IRIS} slit. Two blue dashed lines in panel (b) represent the start and end position of the {\it IRIS} slit. The green dashed line in panel (d) is for the analysis of axial motion of the observed jets.}\label{fig:imaging}
\end{figure*}

\section{FUV and EUV Imaging} \label{imaging}

The emerging negative polarity at the edge of the northern positive polarity of active region NOAA AR12381 studied in \cite{LiuJ2016a} moved further left away from the main active region after 12 UT on July 9th 2015. It experienced some flux cancellation, almost vanished at around 19 UT, merged with other small negative polarities and finally became part of the target negative polarity that we will study in this paper (enclosed in the dashed box in Fig.~\ref{fig:imaging} a).

The online animation of Figure~\ref{fig:imaging} shows jets originating from the target negative polarity from 06 UT to 10 UT on July 10th 2015. Four jets are simultaneously observed by the Solar Dynamics Observatory ({\it SDO}) and Interface Region Imaging Spectrograph ({\it IRIS}) from 07:29 UT to 08:35 UT in this region. Figure~\ref{fig:imaging} shows the direct imaging observation of the fourth jet and its corresponding photospheric line-of-sight magnetic field, by the {\it IRIS} Slit-Jaw Imager \citep[SJI,][panel b and c]{DePontieu2014}, {\it SDO} Atomospheric Imaging Assembly \citep[AIA,][panel d to f]{Lemen2012}, and {\it SDO} Helioseismic and Magnetic Imager \citep[HMI,][panel a]{Schou2012}. 

Slit-jaw images of {\it IRIS} are taken simultaneously in four passbands (C II 1330 \AA{}, Si IV 1400 \AA{}, Mg II k 2796 \AA{}, and Mg II wing 2830 \AA{}) with a spatial resolution of 0.33$''$. These passbands can provide us images from the photosphere (Mg II wing 2830 \AA{}), upper chromosphere (Mg II k 2796 \AA{}), to the transition region (C II 1330 \AA{} and Si IV 1400 \AA{}). These jets show their obvious presence in SJI 1330 \AA{} and 1400 \AA{} FUV, indistinct presence in the SJI 2796 \AA{} NUV and are almost invisible in the other NUV passband, indicating that they should contain materials with temperature at least equivalent to that of the upper chromosphere. Meanwhile, {\it SDO}/AIA provides simultaneous imaging of the jets at seven narrow-band EUV passbands \citep[i.e. Fe XVIII 94 \AA{}, Fe VIII, XXI 131 \AA{}, Fe IX 171 \AA{}, Fe XII, XXIV 193 \AA{}, Fe XIV 211 \AA{}, He II 304 \AA{}, and Fe XVI 335 \AA{};][]{Lemen2012} with a spatial resolution of 1.5$''$. These jets show their existence in all the above passbands, see Figure~\ref{fig:imaging} (d) to (f) as examples at the 304 \AA{} (0.05 MK), 171 \AA{} (0.6 MK) and 335 \AA{} (2.5 MK) passbands. The above observations suggest a temperature ranging from 0.05 MK to at least several MK of the jets' material and their multi-thermal nature. 

\begin{figure*}[tbh]
\centering
\includegraphics[width=\hsize]{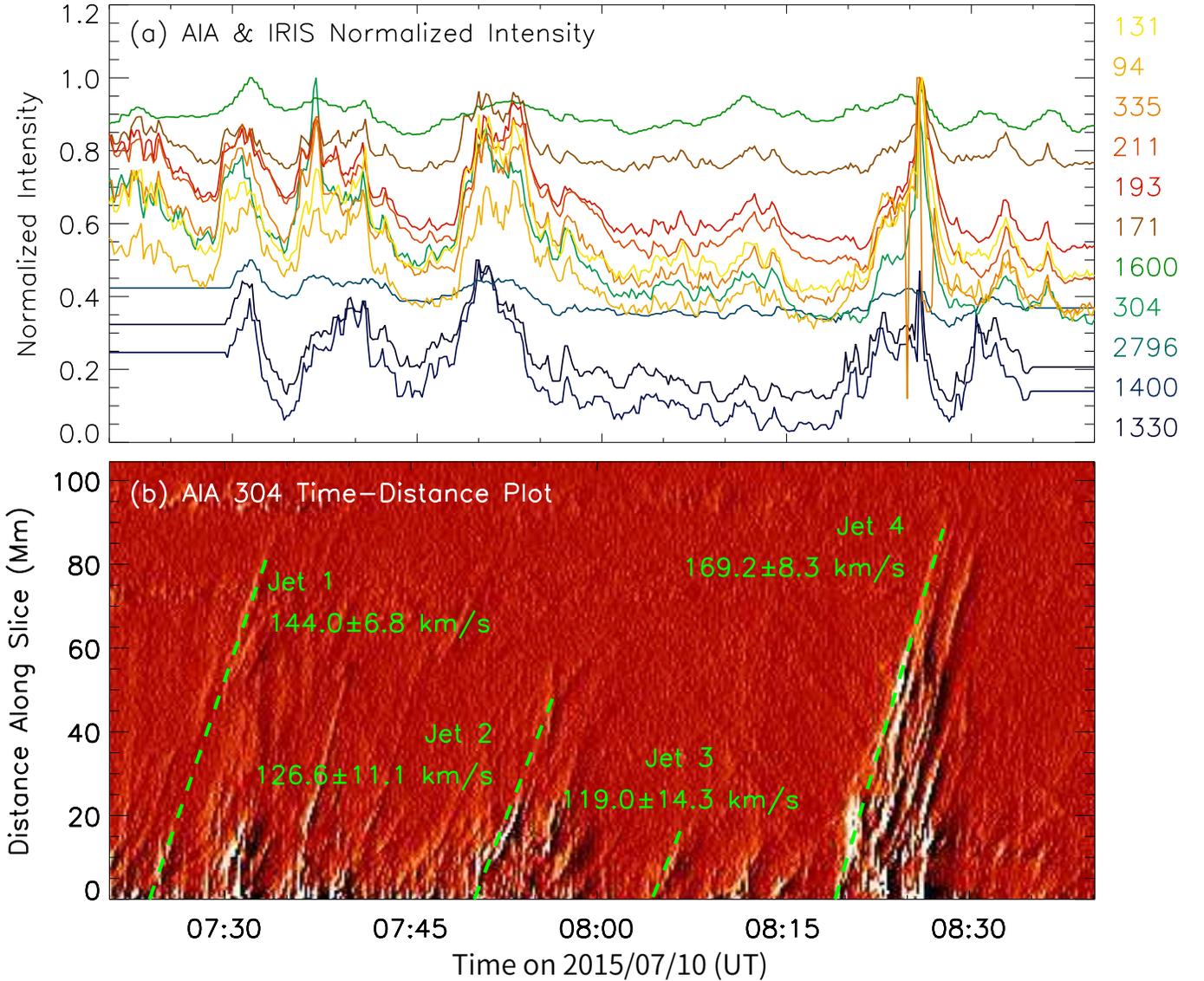}
\caption{Panel (a): Temporal evolution of the normalized average intensity within the black dashed box in Figure~\ref{fig:imaging} at three {\it IRIS}/SJI (shifted 0.5 downward) and eight {\it SDO}/AIA passbands. Panel (b): Running-difference time-distance diagram generated from a 10$''$-wide slit shown as the green dashed line in Fig.~\ref{fig:imaging} (d) at AIA 304 \AA{}. Estimated axial speeds are labeled in green. Errors of speeds are estimated from assuming a 10-pixel uncertainty in the y direction.}\label{fig:curve}
\end{figure*}

Figure~\ref{fig:curve} (a) is the temporal evolution of the normalized average intensity within the black dashed box in Figure~\ref{fig:imaging} (a) at three {\it IRIS}/SJI passbands and eight  {\it SDO}/AIA passbands. It, again, confirms the multi-thermal nature of the observed jets. From a 10-pixel wide slice, shown as the green dashed line in Figure~\ref{fig:imaging} (d), we can generate the corresponding time-distance plots at the above passbands to study the axial behavior of the jets. Figure~\ref{fig:curve} (b) shows one of the generated time-distance plots at the AIA 304 \AA{} passband. All the four jets are labeled by green in chronological order. The four dash lines represent linear fitting results along the jets' trajectories, and their inclination angles indicate the plane of the sky (POS) projected axial speeds ($v_{ap}$) of these jets, ranging from about 120 to 170 km s$^{-1}$.

\section{Spectral Observation} \label{spectral}

From 07:29 UT to 08:35 UT, {\it IRIS} performs a very large dense 192-step raster scan at the edge of the active region NOAA AR12381. It scans a region with a size of 66$''\times$172$''$ from left to right (between the blue dashed lines in Fig.~\ref{fig:imaging} b) for 4 times. Every time, the slit runs through the body of one jet (labeled as Jet 1, Jet 2, Jet 3 and Jet 4 in Fig.~\ref{fig:curve} b). From the online animation, we can find signs of rotational motion of Jet 1, Jet 2 and Jet 4, with Jet 4 revealing the most obvious rotational motion. Jet 3 is too small to resolve its rotational motion if there is any.

The above {\it IRIS} raster scan observation contains several spectral windows, including the C II 1336 \AA{}, Si IV 1403 \AA{} and Mg II k 2796 \AA{}. Because the Mg II k and C II lines are optically thick and have complex line profiles, we use the optically thin line Si IV 1403 \AA{}, which is formed at $log\ T/K = 4.9$ \citep[CHIANTI 7.0,][]{Landi2012}, to explore the LOS Doppler shift signal caused by the jets' rotational motion.

\begin{figure*}[tbh]
\centering
\includegraphics[width=\hsize]{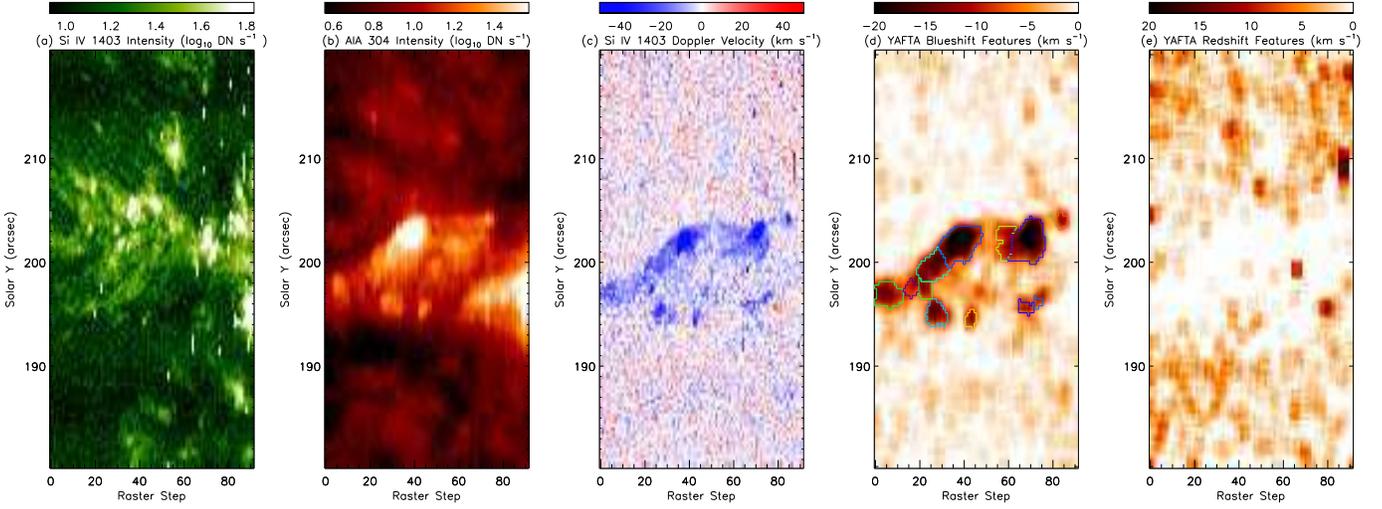}
\caption{Spectral observation of Jet 1 at {\it IRIS} Si IV 1403 \AA{}. Panel (a) and (c): Zeroth and first moments of {\it IRIS} Si IV 1403 \AA{}. Panel (b): Fake raster scan intensity map generated from {\it SDO}/AIA 304 \AA{} observations. Panel (d) and (e): YAFTA feature detection results from the blueshift and redshift part of panel (c), respectively. Colored contours surround the detected features significantly larger than the background noise.}\label{fig:jet1}
\end{figure*}

\begin{figure*}[tbh]
\centering
\includegraphics[width=\hsize]{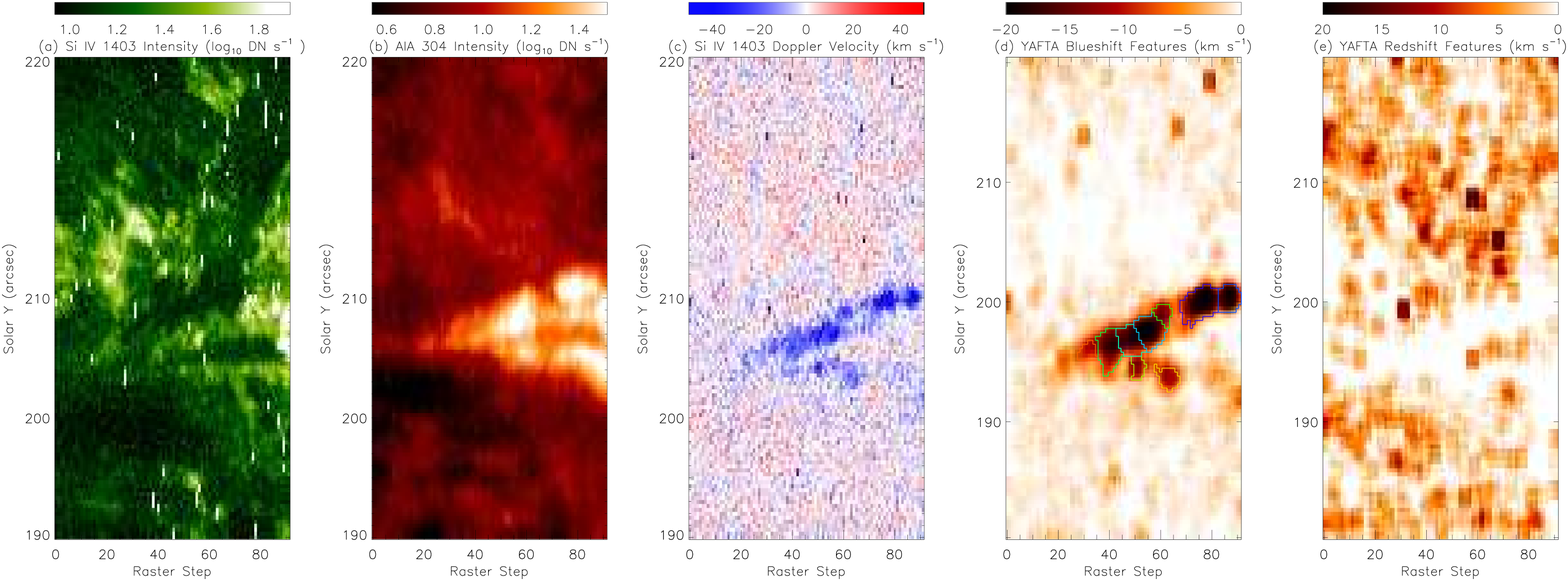}
\caption{Similar to Fig.~\ref{fig:jet1}, but for spectral observation of Jet 2 at {\it IRIS} Si IV 1403 \AA{}.}\label{fig:jet2}
\end{figure*}

Figure~\ref{fig:jet1} shows the spectral observation of Jet 1 at {\it IRIS} Si IV 1403 \AA{}. Figure~\ref{fig:jet1} (a) presents the zeroth moment of the above line, which is derived from \(I_{line} = \int{d(I_{obs}-I_{min})}\), where $I_{obs}$ is the observed spectrograph intensity and $I_{min}$ is the minimum value of $I_{obs}$. It is shown that Jet 1 causes obvious intensity enhancement (bright features from left to right in the middle of Fig.~\ref{fig:jet1} a) at {\it IRIS} Si IV 1403 \AA{}. To compare the features detected in Si IV 1403 \AA{} with imaging observations, we generate a ``fake'' intensity image of Jet 1 at 304 \AA{} (Fig.~\ref{fig:jet1} b), via putting a virtual slit on AIA images (shown as the vertical black solid line in Fig.~\ref{fig:imaging} d) which runs synchronously with the real slit on {\it IRIS}. The generated AIA intensity map gives a similar but clearer intensity response of Jet 1. Please note that, the shape of the jet body shown in these two intensity maps is different from the direct imaging observations, because the intensities in the maps are taken at different instances and the jet keeps evolving.

\begin{figure*}[tbh]
\centering
\includegraphics[width=\hsize]{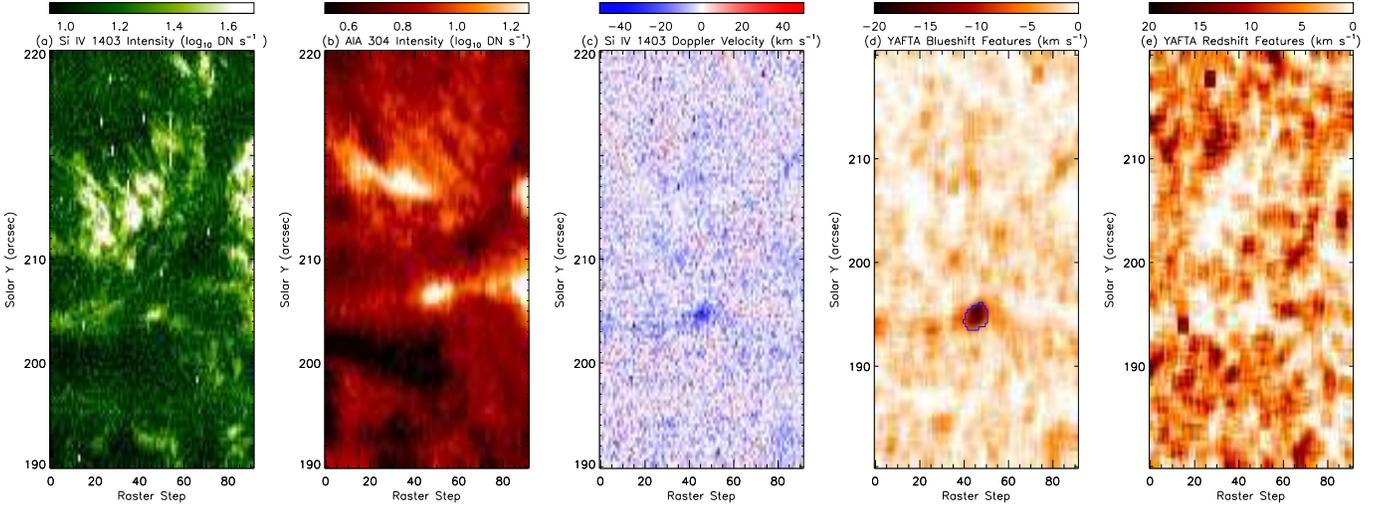}
\caption{Similar to Fig.~\ref{fig:jet1}, but for spectral observation of Jet 3 at {\it IRIS} Si IV 1403 \AA{}.}\label{fig:jet3}
\end{figure*}

\begin{figure*}[tbh]
\centering
\includegraphics[width=\hsize]{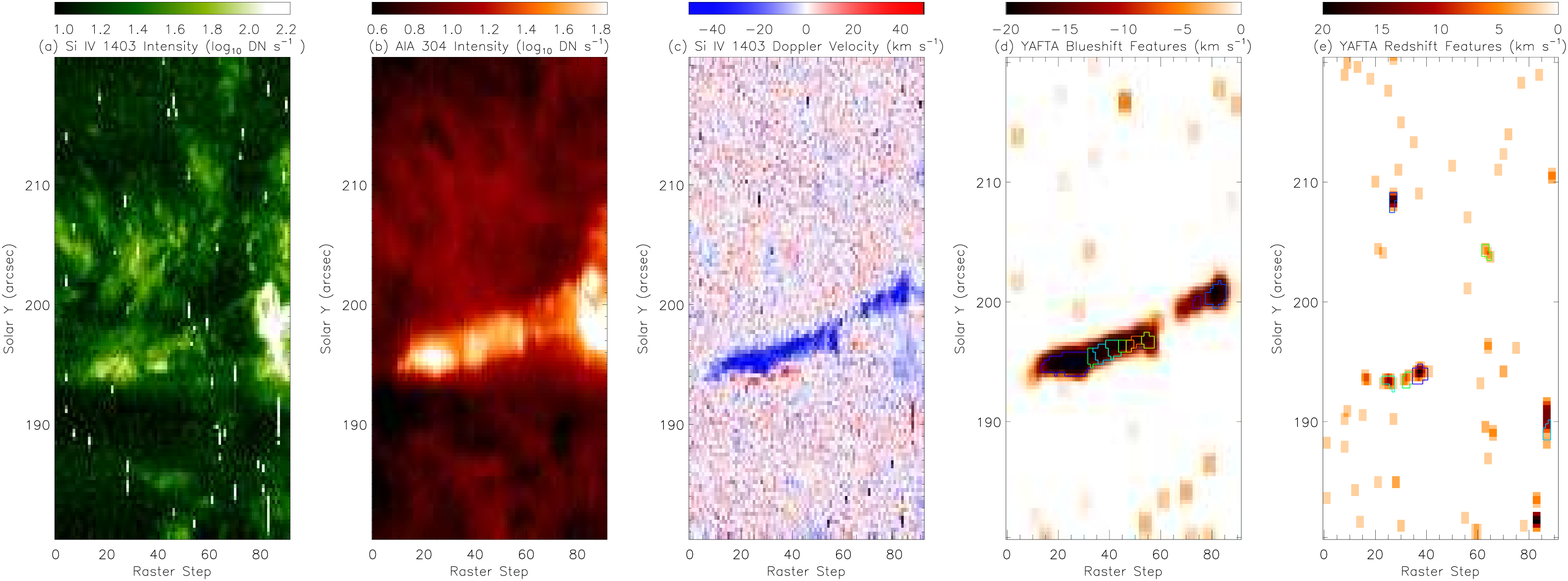}
\caption{Similar to Fig.~\ref{fig:jet1}, but for spectral observation of Jet 4 at {\it IRIS} Si IV 1403 \AA{}.}\label{fig:jet4}
\end{figure*}

The first moment (Doppler shift) of {\it IRIS} Si IV 1403 \AA{}, which is defined by \(v_{dop} = I_{line}^{-1} \int{(I_{obs}-I_{min}})dv_{los}\), is presented in Figure~\ref{fig:jet1} (c). In the above equation, $v_{los}$ is the LOS Doppler velocity, and $I_{obs}$ and $I_{min}$ are the same as previously defined. From a visual check of the Doppler shift map of Jet 1, we conclude that it only has significant blueshift without any redshift. To validate our visual investigation, we divide the Doppler shift map into two parts: one with only blueshift signal and the other with only redshift signal. Then, we apply the Yet Another Feature Tracking Algorithm  \citep[{\it YAFTA},][]{Welsch2003, DeForest2007} to each of the separated maps to recognize signals with values significantly larger than the background noise. The YAFTA method was originally intended to detect different magnetic features on the active region magnetogram, with a flux-ranked, downhill labelling algorithm. Blueshift features detected by the YAFTA method are surrounded by colored contours in Figure~\ref{fig:jet1} (d), in which the darker the color map is the larger value the blueshift presents. Employing this method, we can also estimate the average blueshift of Jet 1, which turns out to be $14.2\pm3.6$ km s$^{-1}$. No redshift features related to Jet 1 are detected, as shown in Figure~\ref{fig:jet1} (e).

Similar investigation is carried out for Jet 2, 3 and 4. In agreement with the results found for Jet 1, Jets 2 and 3 also show significant blueshift without any apparent redshift at {\it IRIS} Si IV 1403 \AA{} (Fig.~\ref{fig:jet2} and~\ref{fig:jet3}). The average blueshift of Jet 2 and 3 determined by the YAFTA method is $15.5\pm3.2$ km s$^{-1}$ and $\sim 11.2$ km s$^{-1}$, respectively.  The overall blueshift patterns of these three jets suggest that they are all inclined out of the POS. However Jets 1 and 2 both show clear rotational patterns in imaging observations (see the online animation). The reason why they do not cause clear redshift signals in the Doppler shift maps might be that: (1) they have not started rotating while the {\it IRIS} slit passes by, (2) their rotational motion is not fast enough to counteract the blueshift caused by the LOS projection of their axial motion, so that the overall Doppler velocity still only shows blueshift.

\begin{table*} \small
\centering
\linespread{1.5} \caption{Parameters of 4 homologous jets deduced from imaging and spectral observations}

\begin{tabular}{c|c|c|c|c|c|c}
\hline
Jet NO. & Onset Time  & Axial Speed & Blueshift & Redshift & Rotational Speed \\
 & (UT) & (km/s) & (km/s) & (km/s) & (km/s) \\
\hline
1 & 07:23 &144.0$\pm$6.8 & 14.2$\pm$3.6 & -- & -- \\
2 & 07:49 & 126.6$\pm$11.1 & 15.5$\pm$3.2 & -- & -- \\
3 & 08:04 & 119.0$\pm$14.3 & $\sim$11.2 & -- & -- \\
4 & 08:18 & 169.2$\pm$8.3 & 28.6$\pm$3.0 & 7.3$\pm$2.1 & 28.2$\pm$4.0 \\
\hline
\end{tabular}
\label{tb1}
\\
Note. ``Axial Speed" is the POS projection of jets' axial speed.
\end{table*}

Differently, Jet 4 is undergoing very clear rotational motion when the {\it IRIS} slit passes by. This can also be verified by the spectral observations. The rotational motion causes both blueshift (Fig.~\ref{fig:jet4} d) and redshift (Fig.~\ref{fig:jet4} e) at Si IV 1403 \AA{}. From Figure~\ref{fig:jet4} (c), we find that the main body of Jet 4 shows stronger blueshift than the previous three jets, while the lower edge of it shows significant redshift. This suggests that the rotational direction of Jet 4 is left-handed. The average blueshift and redshift velocities detected are $28.6\pm3.0$ km s$^{-1}$ and $7.3\pm2.1$ km s$^{-1}$, respectively. Both the blueshift and redshift detected above are the average values of the combined effect of the rotational motion and the LOS axial motion of the jet. 

\begin{figure*}[tbh]
\centering
\includegraphics[width=\hsize]{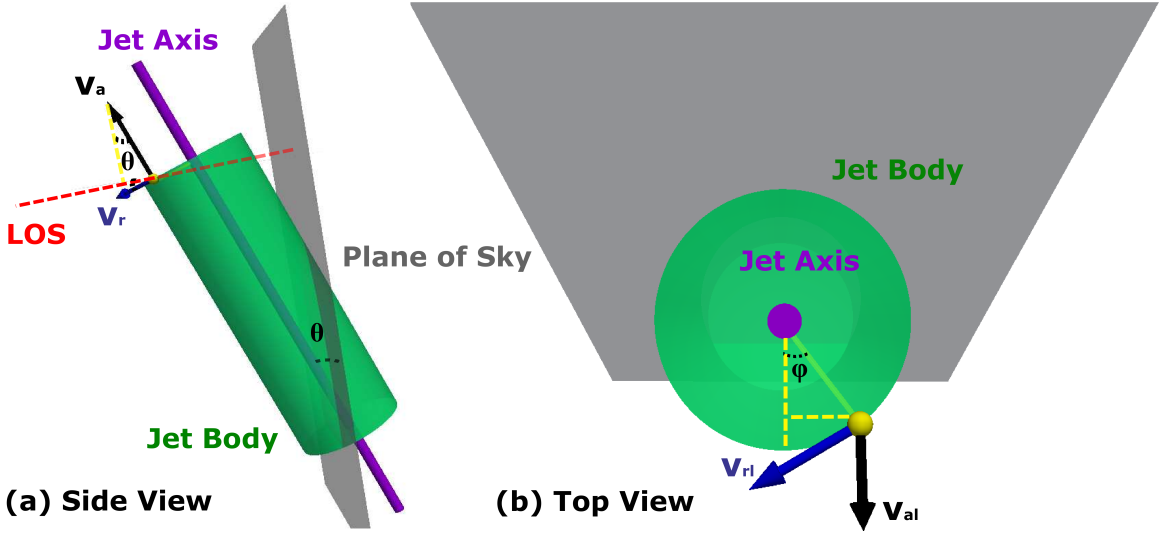}
\caption{Panel (a): Side view of the model employed to estimated the rotational speed of the jet. Panel (b): Top view of the model. Green cylinder represents the jet body, purple line its axis, yellow sphere a point at its surface, red line the LOS and the gray shadow the POS. $v_{al}=v_a\sin{\theta}$, is the the LOS projected axial speed of the jet, where $v_{a}$ is its axial speed and $\theta$ the inclination angle of its axis relative to the POS. $v_{rl}=v_r\cos{\theta}$ with $v_r$ the rotational speed. Scales do not correspond to real values.}\label{fig:mod}
\end{figure*}

Assume the jet has as a cylindrical shape and rotates as a solid body, which can be found appropriate from the online animation and observations of other coronal jets \citep[e.g.,][]{LiuJ2014}. Figure~\ref{fig:mod} shows a toy model (with panel (a) side view and panel (b) top view) of how the movement of a point (yellow sphere) at the jet's surface contributes to the detected LOS velocity. This point, with an angle of $\varphi$ from 0 to $\frac{\pi}{2}$ either clockwise or counter-clockwise from the vertical dashed yellow line in Figure~\ref{fig:mod} (b), has a LOS velocity of $v_{a}\sin{\theta}+v_{r}\cos{\theta}\sin{\varphi}$ (clockwise from the vertical dashed yellow line) or $v_{a}\sin{\theta}-v_{r}\cos{\theta}\sin{\varphi}$ (counter-clockwise from the vertical dashed yellow line), where $v_{a}$ is the axial speed of the jet, $v_{r}$ the rotational speed of the jet, and $\theta$ the inclination angle of the jet axis relative to the POS. Thus we have the following two equations:

\begin{equation} \label{eq1}
\int_0^{\frac{\pi}{2}}{v_{a}\sin{\theta}-v_{r}\cos{\theta}\sin{\varphi}d\varphi} = -\frac{\pi}{2}\cdot v_{blue}
\end{equation}

\begin{equation} \label{eq2}
\int_0^\frac{\pi}{2}{v_{a}\sin{\theta}+v_{r}\cos{\theta}\sin{\varphi}d\varphi} = \frac{\pi}{2}\cdot v_{red}.
\end{equation}

Here, $v_{blue}$ ($v_{red}$) is the average blueshift (redshift) detected above. Define $v_{al}=v_a\sin{\theta}$ and $v_{rl}=v_r\cos{\theta}$, which is the projected axial and rotational speed in the plane perpendicular to the POS respectively, we can obtain $v_{rl}$ as $28.2\pm4.0$ km s$^{-1}$ and $v_{al}$ as $-10.6\pm2.6$ km s$^{-1}$.

Knowing the LOS ($v_{al}$) and POS ($v_{ap}$) projected axial speeds of all 4 jets, one may consider to estimate the inclination angles ($\theta$) of their axes relative to the POS employing triangular functions ($\theta=\arctan{\frac{v_{al}}{v_{ap}}}$), and further derive the real value of the rotational speed $v_r$ from $v_r=v_{rl}/\cos{\theta}$. Unfortunately, this could sometimes be problematic. Because it is usually hard to tell whether the POS projected axial speeds estimated by imaging observations (Fig.~\ref{fig:curve}) are bulk or wave speeds (while LOS projected axial speeds estimated by spectral observations should be bulk speeds). 3D numerical simulations by \cite{Lee2015} and \cite{Pariat2016} have shown that in coronal conditions the bulk flow of the jet could be disconnected and very different from the compression front of the torsional wave which is generated during the magnetic reconnection. If we assume the POS projected axial speeds ($v_{ap}$) estimated in Table~\ref{tb1} are bulk speeds, we can infer the inclination angles ($\theta$) of their axes ranging from $4^\circ - 7^\circ$, suggesting almost horizontal field lines originating from their source region. This is inconsistent with what we will show in Figure~\ref{fig:mag} (a). The open field lines (blue) derived from non-linear force free extrapolation in Figure~\ref{fig:mag} (a) have much larger inclination angles from 30$^\circ$ to 45$^\circ$. However, because the real magnetic field has not to be necessarily force-free, the inclination angle derived above is not ready to be used either, and may cause large error. Considering the above difficulties to derive the real rotational speed of the jet, we tend to use its LOS component as an approximation in the rest of the paper.

All the results, including the apparent onset time, the POS axial speed, blue and redshift velocities,  inclination angle and rotational speed of all 4 jets deduced from both imaging and spectral observations, are shown comprehensively in Table~\ref{tb1}.

\section{The Twist} \label{magnetic}
Taking the rotational speed of the fourth jet $v_{r}\sim28.2$ km s$^{-1}$ and its average width $d\sim$4.9 Mm into account, the estimated rotational period is $P=\pi d/(v_{r})\sim$546 s. From {\it SDO}/AIA imaging observations, as shown in the online animation, we find that the rotational motion of Jet 4 lasts for $T\sim$696 s from around 08:19 UT. Then, we estimate the total number of turns Jet 4 rotates: $T/P\sim$1.3. As described in the third paragraph in Sect.~\ref{intro}, if the rotational motion of the jet is associated with the unwinding process after magnetic reconnection, Jet 4 should have released a twist of $\sim2\pi\times1.3=2.6\pi$ in total. Considering that we have underestimated the rotational speed of the jet as demonstrated in the previous section, the twist angle estimated here should be a lower limit of the real value.

The above results suggest the possibility of a flux rope with twist angle at least 2.6$\pi$ radians existing before the eruption of Jet 4. To examine this, we conduct a non-linear force-free field extrapolation \citep[NLFFF,][]{Wiegelmann2008} from the vector magnetic field of the whole active region, obtained from its corresponding Space-Weather HMI Active Region Patch 5745 \citep[SHARP,][]{Hoeksema2014}. To meet the force-free and torque-free condition, the observed photospheric vector magnetic field has to be pre-processed in the above NLFFF. The magnetic field data is smoothed (the factor $\mu_4$ of the smoothing term in Eq.6 in \cite{Wiegelmann2006} is set to 0.01 in our extrapolation) and mapped to chromosphere where the actual extrapolation starts \citep{Wiegelmann2006}, and the resolution is downgraded by a factor of 4 to reduce the computation. The same method, which introduces two parameters ($\sigma_J$: sine of the current-weighted average angle between the current $\mathbf{J}$ and magnetic field $\mathbf{B}$; and $<|f_i|>$ the average fractional change of flux), as described in Eq.13 and Eq.15 of \cite{Wheatland2000}, is employed to evaluate whether the extrapolated magnetic field meets the force-free and divergence-free conditions. For a perfectly force-free and divergence-free field, $\sigma_J$ and $<|f_i|>$ would both be 0. In this case, $\sigma_J$ and $<|f_i|>$ are 0.23 and 8.3$\times 10^{-3}$, respectively, suggesting a fairly good extrapolation result.

\begin{figure*}[tbh]
\centering
\includegraphics[width=0.9\hsize]{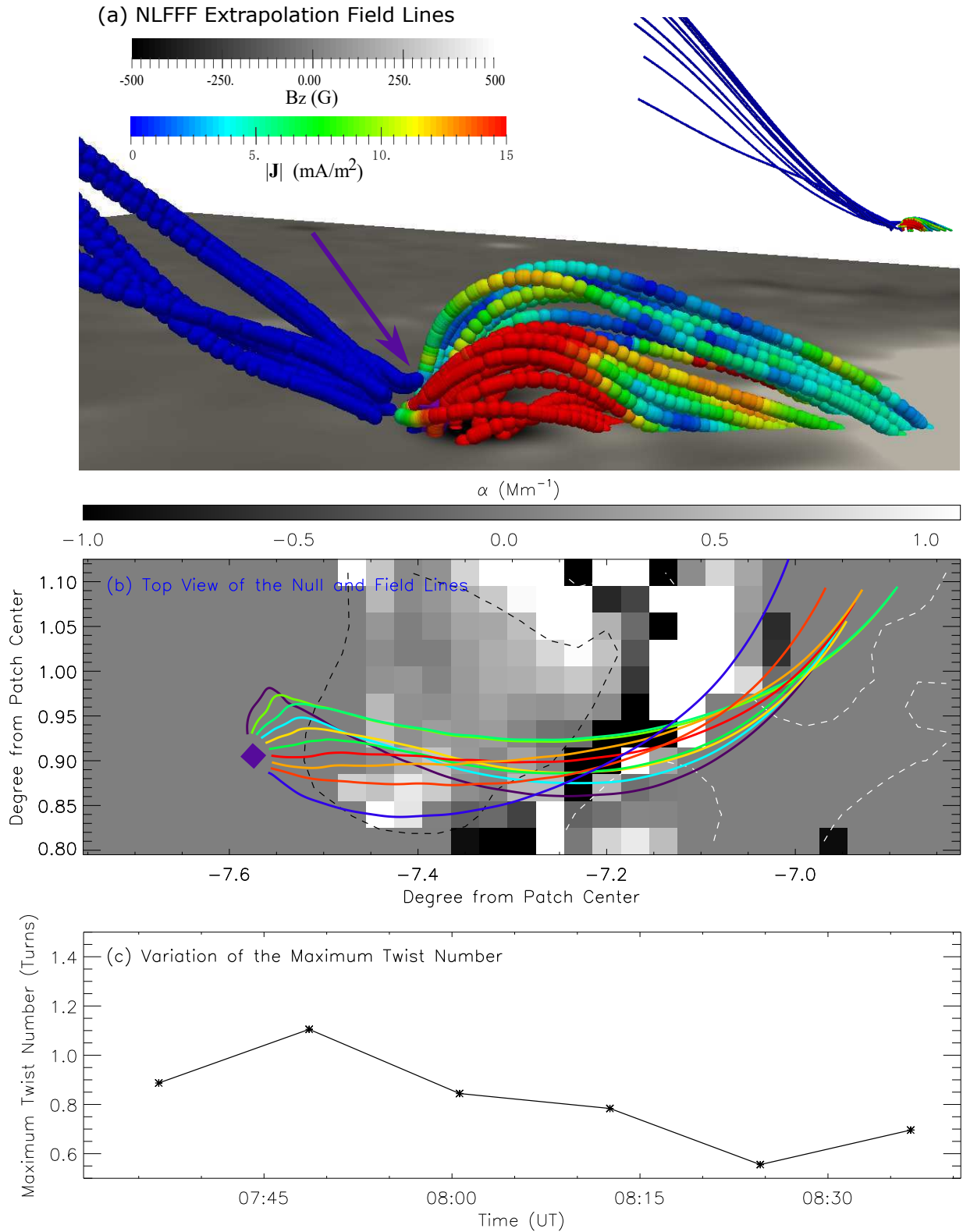}
\caption{Panel (a): Magnetic field by NLFFF extrapolation at 08:12 UT. To make the low-embedded high-current loops visible, the pixel size in z-direction has been magnified by a factor of 2. Single-color blue lines are open field lines. Colored lines represent closed field lines with colors indicating modulus of current density $\mathbf{J}$. The purple sign indicated by the purple arrow is at the detected null point. Embedded figure on the top-right of this panel gives a side view of the extrapolated magnetic field lines with a larger FOV extending to the left edge of the active region patch with original z-direction scale. Panel (b): the photospheric force-free parameter $\alpha$ (white-black background), the vertical magnetic field (black and white dashed contours), the null point (purple diamond) and the twisted magnetic field lines (colored lines) at 08:12 UT. Warmer colors stand for higher twists of the field lines. Panel (c): Variation of the maximum twist number of magnetic field lines associated with the detected null point.}\label{fig:mag}
\end{figure*}

Figure~\ref{fig:mag} (a) shows the extrapolated field lines nearby the source region of the jet (negative polarity enclosed in the dashed box in Fig.~\ref{fig:imaging} a), at around 08:12 UT, right before the eruption of Jet 4. Single-color blue lines on the left indicate (locally) open field lines, with colored lines on the right closed loops. Colors of the closed loops indicate the local current density $|\mathbf{J}|$, with warmer color corresponding to larger current density. Among these loops, we can find several lower embedded ones, with high current, moderately winding around each other and seemingly form a flux rope.

Following \cite{Haynes2007} and \cite{LiuR2016b}, we solve the $\mathbf{B} = 0$ condition with the iterative Newton-Raphson method to locate null positions to subgrid precision. Totally, 25 null points are found in the entire active region from NLFFF extrapolation. However, only one null point is found in the vicinity (a box sized 135$''\times$115$''$) of the jet's source region. The location of the null point is shown as a purple sign indicated by the purple arrow in Figure~\ref{fig:mag} (a). The height of the null point is about 300 km (pixel size in the extrapolated field is $\sim$360 km), indicating that it is low located in the solar atmosphere. This null point is located between the closed and open fields with high current density (red color) around it. The above topology of magnetic field is highly in favor of magnetic reconnection and then the eruption of jets. We should notice that, in practice, one may find more photospheric null points in an extrapolated (force-free or potential) field based on original magnetograms than the current NLFFF we employ that is based on pre-processed magnetograms. The above null point could only be important if it is also unique in another different extrapolation and stable existing at different times. Thus, we perform the same null-point detection method to a potentially extrapolated field. The uniqueness of the above null point in the potential field extrapolation is then proved by the fact that it is again the only null point detected in the vicinity of the jet's source region within a 135$''\times$115$''$ box. To test its stable existence, we detect all null points in NLFFF extrapolated fields from 07:36 UT to 08:36 UT. The above null point is found to be present at the same location in all the extrapolated fields.

Figure~\ref{fig:mag} (b) presents the top view of the photospheric force-free parameter $\alpha$ (white-black image, $J_z/B_z$), the vertical magnetic field (black and white dashed contours at $\mp$150 G), the detected null point (purple diamond) and the twisted magnetic field lines (colored lines) associated with the null point, at 08:12 UT. Warmer color of the field line indicates higher twist number $T_w$. $T_w$ (in units of turns) is calculated by the following equation:

\begin{equation}\label{eq3}
T_w = \frac{1}{4\pi}\int_L{\alpha dl}\ ,
\end{equation}

\noindent which is the integration of the force-free parameter $\alpha$ along the field line and measures how much two neighboring field lines twist each other \citep{Berger2006}. The twisted field lines shown in Figure~\ref{fig:mag} (b) all have $T_w$ above 0.6 and a maximum $T_w$ of about 0.8 turns. For finding and tracking the field lines associated with the null point accurately, we used a subgrid step (0.05 pixel) above. A side-effect of the highly accurate tracking is those local wiggles which could be found around the eastern ends of the field lines in Figure~\ref{fig:mag} (b). To examine how much these wiggles contribute to the twist number $T_w$ estimated above, we track these field lines again at a larger step (2 pixels). Field lines tracked by the large-step scheme are found to be smooth without wiggles. The twist numbers of the field lines obtained from small- and large-step schemes have differences ranging from -9\% to 14\%, indicating a negligible contribution of these wiggles to the twist number estimated above. The opposite sign of the twist the pre-eruption twisted fields contain (positive) and the direction the jet rotates (left-handed), is consistent with the scenario that the rotational motion of the jet is a result of the untwisting process after magnetic reconnection. The above result is also consistent with that the photospheric $\alpha$-map at the jets' footpoint region (black dashed contour in Fig~\ref{fig:mag} b) is dominated with positive values (with average $\sim$0.2 Mm$^{-1}$).

The field line with the maximum twist is colored by red in Figure~\ref{fig:mag} (b), which has a length of 9.6 Mm. Figure~\ref{fig:mag} (c) depicts the variation of the maximum twist number obtained from magnetic field lines associated with the detected null point. It shows an overall decrease of the twist the jets' source region contains with the erupting of jets. Detailed one-to-one relationship between the observed jets and the derived twist numbers is impossible to be carried out due to the low cadence (12 min) of the vector magnetic field. 

Before comparing the twist number a jet releases with that its source region flux rope contains, we should pay particular attention to the following issues: (1) twist is not always a conserved quantity, and (2) the difference between static twist and dynamic twist. Jets usually contain many strands which may wind around each other during the eruption. If we take a snapshot during the eruption and count how many turns these strands wind around each other, the resulting static twist could be very different with the twist contained in the pre-eruption flux rope. The dynamic twist represents how many turns a coronal jet rotates during its eruption. The above two issues make the relationship between the dynamic twist and the twist the jet's source region contains complex. Thus, numerical simulations are needed. \cite{Pariat2016} performed a series of 3D numerical simulations of solar jets in conditions with different plasma $\beta$. It is found the dynamic twist is almost the same with that injected into the system before eruption when plasma $\beta$ is less than 1 - corresponding to coronal conditions. Then, we can make a careful conclusion that from the number of turns Jet 4 performs, the pre-eruption flux rope at its source region should have contained a twist number of at least 1.3. However, the maximum twist number (0.8) derived from the NLFFF extrapolation is less. This could be caused by the following: (1) $T_w$ defined here is usually less than the traditional defined twist \citep[see, e.g.,][]{LiuR2016, WangY2016}. (2) The real magnetic field in the solar atmosphere, especially at sites with high current, is usually very complex and not necessarily force-free. (3) When we calculate the $\alpha$ at each point along a traced magnetic field line, we use its neighboring points which could belong to other field lines to calculate the derivations of the magnetic field. This could also decrease the estimated maximum twist number. Future study on the magnetic field with higher resolution/cadence and non-force-free field extrapolation method might give higher $T_w$ consistent twist numbers and cleared evidences of the flux rope formed by the twisted field lines.

\begin{figure*}[tbh!]
\centering
\includegraphics[width=\hsize]{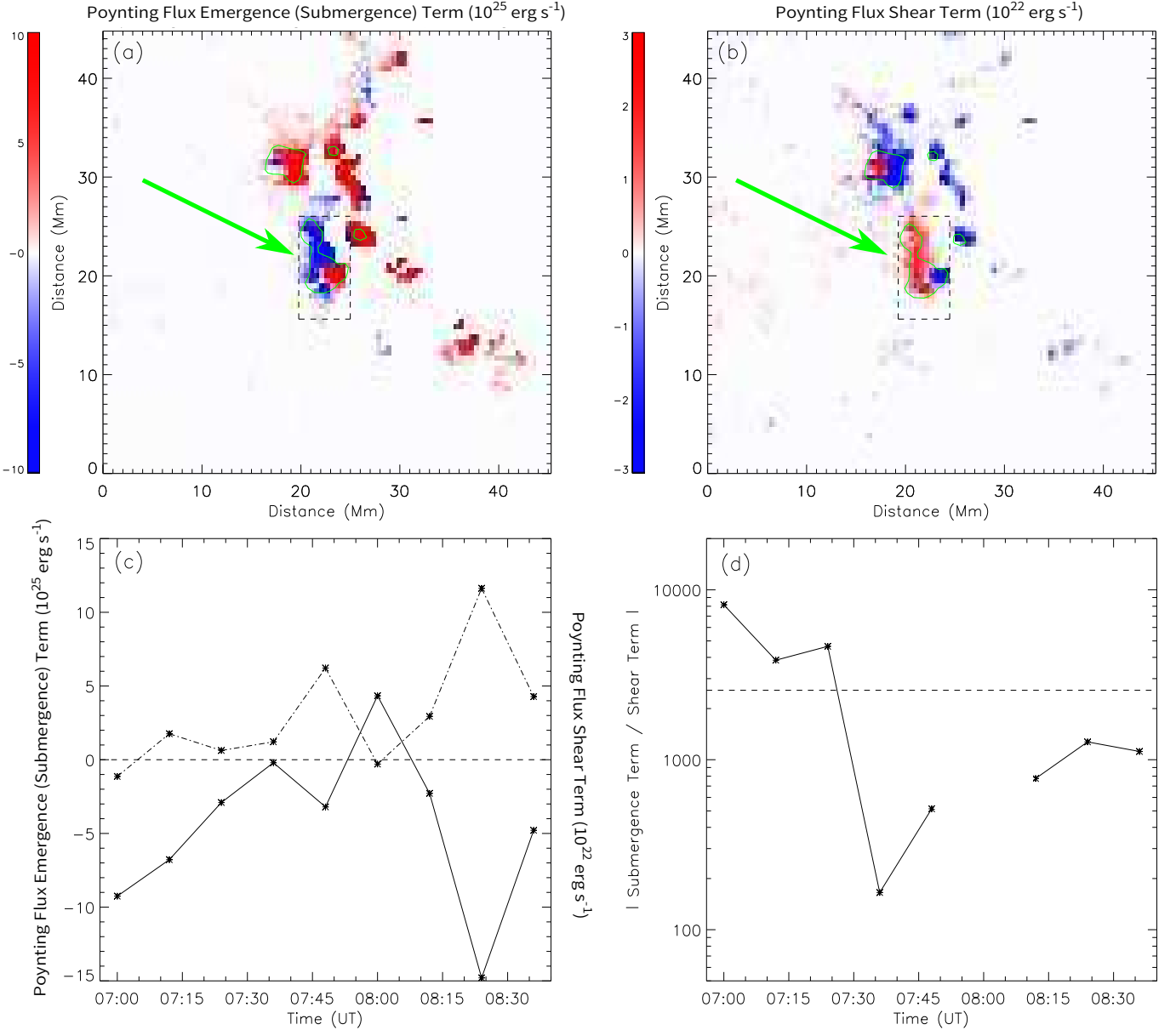}
\caption{Panel (a) and (b): Emergence (submergence) and shear terms of photospheric Poynting flux at 08:12 UT. To highlight the distribution at the source region of jets (indicated by the green arrow), Poynting fluxes at the main positive polarity are set to be zero (blank). Green contours in both panels indicate a vertical magnetic field level of -150 G. Panel (c): Variation of the average emergence or submergence (positive or negative, solid curve) and shear (dash-dotted curve) terms of the Poynting flux in the region enclosed by the dashed box in panel (a) and (b). Solid curve in panel (d) shows the modulus of the ratio between the mean submergence and shear terms within the above dashed box. The point at 08:00 has been omitted because the Poynting flux shows emergence other than submergence at that time. The horizontal dashed line is the average ratio $\sim$2560.}\label{fig:poynting}
\end{figure*}

Highly twisted magnetic flux ropes are subjects to kink instability.  There are several thresholds based on theoretical analyses. The Kruskal-Shafranov limit \citep{Kruskal1958, Shafranov1957} gives a threshold 2$\pi$ of the total twist angle for the onset of kink instability in axisymmetric toroidal magnetized plasma columns. Further study on force-free coronal loops with uniform twist suggests a maximum twist angle of 2.5$\pi$ a kink-stable, cylindrical flux tube might contain \citep{Hood1981}. 3D MHD numerical simulations on the generation of solar jets \citep{Pariat2009} give a slightly higher limit of the twist angle 2.6$\pi$ injected into the system for the onset of kink instability and the eruption of a jet. Contrast to these fixed threshold values, \cite{Dungey1954} suggested that a magnetic flux rope can contain a total twist angle of 2$l/R$ before it becomes kink unstable, where $l$ and $R$ are the length and radius of the flux rope, respectively. Their theoretical result was recently confirmed by \cite{WangY2016} with 115 magnetic clouds observed at 1 AU. We take the radius of the fourth jet ($\sim$2.45 Mm) and the length of the field line ($\sim$9.6 Mm, red one in Fig.~\ref{fig:mag} b) which has the maximum twist, as approximations of the radius and length of the flux rope. It gives a twist angle of $\sim$2.5$\pi$. The twist released by Jet 4 ($\sim$2.6$\pi$), estimated in this paper, is consistent with the above thresholds, suggesting there might be kink instability happening before the eruption of this jet. However, this should be further examined by more direct observations of the kink instability.

\section{Magnetic Flux Cancellation} \label{cancel}

\begin{figure*}[tbh]
\centering
\includegraphics[width=\hsize]{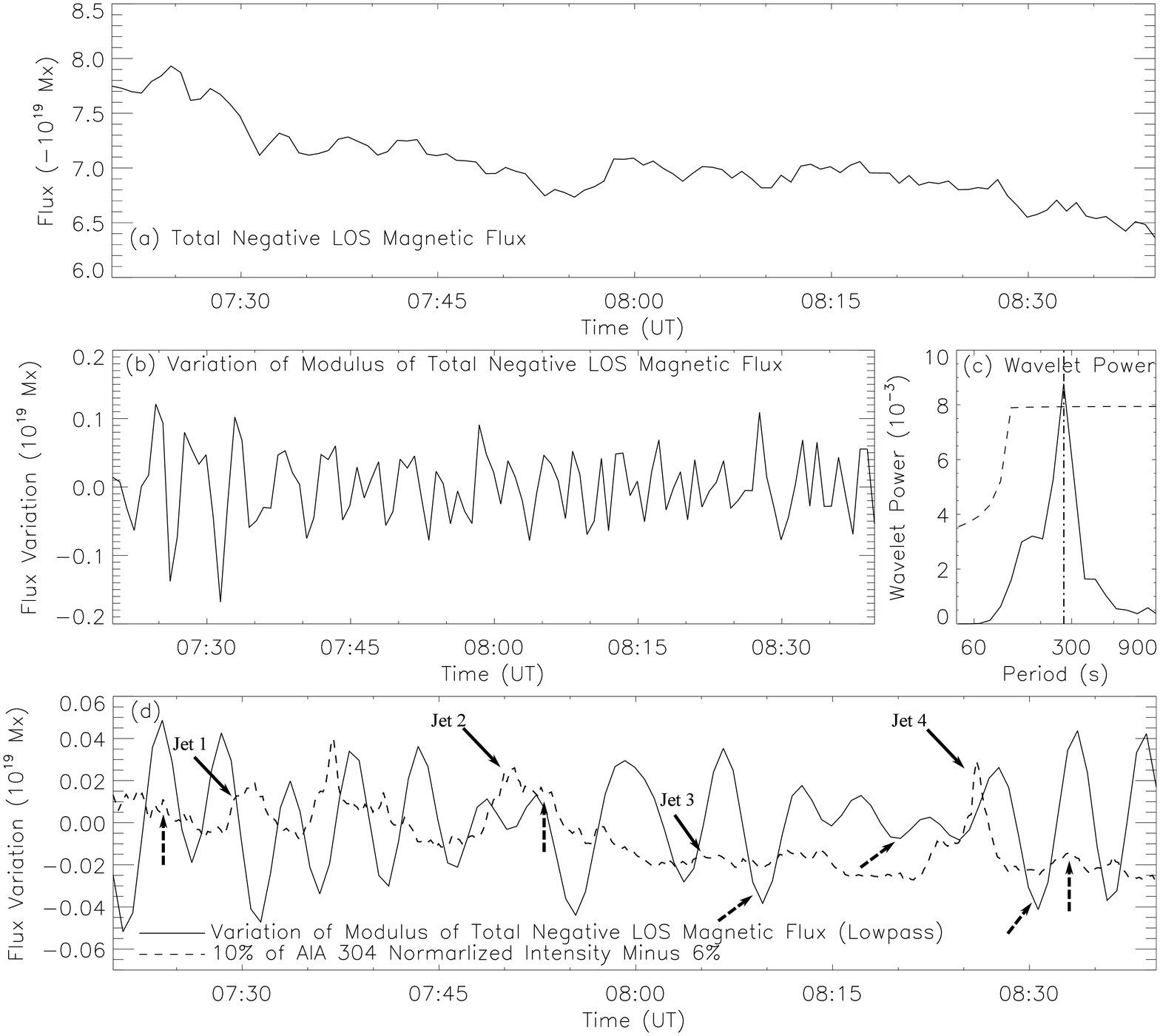}
\caption{Panel (a): Temporal variation of the total LOS negative flux at the source region of jets (black dashed box in Fig.~\ref{fig:imaging} a). Panel (b) and (c): Temporal variation of the modulus of the total negative flux after removing the unitary trend of decreasing, and its corresponding wavelet power. Dashed line in panel (c) indicates a significance level of 95\%. Vertical dash-dotted line coincides with the peak period. Panel (d): solid curve is the low-bandpass result of panel (b). Dashed curve is generated via taking 10\% of the AIA 304 \AA{} normalized intensity within the black dashed box in Fig.~\ref{fig:imaging} (a) and shifting 0.06 downward.}\label{fig:flux}
\end{figure*}

In this section, we will study the relationship between the occurrence of the observed jets and the photospheric magnetic field variations to investigate how the magnetic field and energy flow influences the eruption of the observed jets. The Poynting flux across the photospheric boundary can be estimated by \citep{Kusano2002, LiuR2016}: 

\begin{equation}
\label{eq4}
\frac{dP}{•dt}\Bigg{|}_S=\frac{1}{4\pi}\int_S B_t^2 V_{\bot n}\ dS - \frac{1}{4\pi}\int_S 
(\mathbf{B_t} \cdot \mathbf{V_{\bot t}})B_n\ dS
\end{equation}

\noindent where, $n$ and $t$ refer to the normal and tangential directions, respectively. The vector magnetic field $\mathbf{B}$ is from the SHARP data \citep{Sun2013}. To obtain the photospheric vector velocity field, we apply the Differential Affine Velocity Estimator for Vector Magnetograms \citep[DAVE4VM,][]{Schuck2008} with a window size of 19 pixels following \cite{LiuY2014} to the time series of the vector magnetograms. The first and second terms in the equation are the vertical (emergence/submergence) and horizontal (shear) components, respectively.

Figure~\ref{fig:poynting} shows the distribution of the Poynting flux emergence (submergence) and shear term in panel (a) and (b) respectively, at 08:12 UT before Jet 4 erupts. Green contours in both panels indicate a vertical magnetic field level of -150 G. Red (blue) colors stand for positive (negative) values of the Poynting flux. Green arrows in both panel denote the location of the jets' source region. We note that the distribution of enhanced Poynting flux introduced by either emergence (submergence) or shear matches the jets' source region very well. Even though both positive and negative values can be found at the source region in panel (a), it is dominated by negative ones, which means significant submergence process. A temporal variation of the average emergence/submergence (solid curve) and shear (dash-dotted curve) Poynting flux in the region enclosed by the dashed black box in panel (a) and (b) is shown in Figure~\ref{fig:poynting} (c). From 07:00 to 08:36 UT, 8 frames of Poynting flux distribution are obtained. Except 1 instance at 08:00 UT, all the estimates show much stronger submergence (negative) than emergence (positive). On the other hand, a temporal variation of the absolute value of the submergence term over the shear term at the jets' source region is shown in panel (d), suggesting that the submergence term is always 2 to 4 orders of magnitude larger than the shear term. Considering the close relationship between photospheric submergence and magnetic cancellation \citep[e.g.,][]{Chae2004}, we suspect that these jets could be related to the magnetic cancellation at their source region.

The black solid curve in Figure~\ref{fig:flux} (a) describes how the total LOS negative flux of the jets' source region (enclosed in the dashed box in Figure~\ref{fig:imaging} a) varies during the period from 07:20 to 08:40 UT when there is no visible negative magnetic flux moving out of the box, allowing us to further examine our above conjecture. It is shown that the total negative flux decreases from $\sim-7.7\times10^{19}$ Mx to $\sim-6.6\times10^{19}$ Mx. For a period of 80 min, the solar rotational effect can only causes a 0.008\% reduction of the observed LOS magnetic flux. At the mean time, we can also find some local minimums superimposed on the overall trend of decreasing.

After removing the unitary trend of decreasing, we obtain a local variation of the modulus of the total negative magnetic flux shown in Figure~\ref{fig:flux} (b). Applying a wavelet analysis to the variation, we can find a dominant period of $4.4\pm1.0$ min (Fig.~\ref{fig:flux} c), which is consistent with that of the solar {\it p}-mode oscillation (3 to 5 min). To remove this high-frequency component in the variation, we apply a low-bandpass filter (removing variation with period less than 4.5 min) and the resulted variation is shown as the solid curve in Figure~\ref{fig:flux} (d). The variation of the normalized intensity at AIA 304 \AA{} at the jet's source region is also shown (dashed curve) in the diagram, with peaks denoting eruption of jets. As expected, almost all jets correspond to local minimums in time, which again confirms our previous speculation from studying the photospheric Poynting flux, that these observed jets are related to the magnetic flux cancellation at their source region. 

We also notice that there are (1) few 304 \AA{} brightness enhancements (indicated by vertical dashed arrows) not corresponding to local minimums and (2) few local minimums (indicated by inclined dashed arrows) not corresponding to apparent jet eruptions. A review on the observations of solar X-ray and EUV jets by \cite{Innes2016} has shown that although magnetic flux cancellation could be a dominant process during jet eruptions, it does not always happen (cf. Table 1 in their paper). Limited temporal and spacial resolution, and artificial effects including smoothing, edge effects and band-filter window size selection, could all lead to the absence of cancellation during jet eruptions in observations. On the other hand, similar with the difference between confined and eruptive flares, we cannot expect all cancellations correspond to apparent jet eruptions, due to different local magnetic field configurations which also evolves with time. Besides, as found in \cite{Pariat2016}, straight jet phase before helical jet phase usually correspond to less obvious jets in coronal conditions.

\section{Summary} \label{summary}
In this paper, we performed a detailed study on the imaging and spectral observations of 4 homologous jets and their photospheric conditions. We summarize as follows:

Simultaneous imaging observations from {\it SDO}/AIA and {\it IRIS}/SJI reveal the multi-thermal nature of the observed jets with temperatures ranging from 0.05 MK to at least several MK. The POS projected axial speeds of these jets are found to range from 120 to 170 km s$^{-1}$. Employing the {\it IRIS} raster scan spectral data at Si IV 1403 \AA{}, we identify clear response of these jets at the temperature of 0.08 MK. First moment of the above line shows only blueshift signals resulted from the first three jets, indicating that they are inclined out of the POS. Both blueshift and redshift signals are detected corresponding to the fourth jet, suggesting its obvious rotational motion with average speed at least 28.2 km s$^{-1}$. 

Knowing the rotational speed of the fourth jet, we then infer it rotates about 1.3 turns during its lifetime - which indicates an associated twist number of 1.3 turns having been released. This value is consistent with theoretical kink-unstable thresholds and suggests the existence of a flux rope with a twist number of at least 1.3 turns before the eruption of this jet. Employing NLFFF extrapolation, we identify a bunch of twisted magnetic field lines with high current density, and the existence of a null point between the twisted and open field lines. The jet is found to rotate in the opposite sense of handness to the photospheric $\alpha$ and the twist of the field lines, confirming the unwinding of the twist by the jet's rotational motion. The maximum twist number is estimated to be around 0.8 turns, which is lower than the twist the jet releases, indicating that the employed methods in this paper should have underestimated the twist number.

Investigation of the Poynting flux across the photosphere using the vector magnetic field data at the jets' source region shows both enhanced submergence and shear terms of the Poynting flux. However, the shear term is generally 2 to 4 orders of magnitude lower than the submergence. The above results suggest the high possibility of magnetic flux cancellation taking place at the jets' source region. The temporal evolution of the total negative flux at the jets' source region confirms the magnetic cancellation. The nearly one-to-one correspondence between the local minimum of the flux and the eruption of jets, validates the close relationship between the photospheric magnetic cancellation and the jets.

To conclude, we have identified: (1) magnetic twist a jet releases and that it is consistent with theoretical twist-unstable thresholds; (2) the existence of twisted field lines and associated null point at the jet's source region; (3) the unwinding of the twist by the rotational motion of the observed jet; and (4) evidences of photospheric magnetic flux cancellation during the eruptions. Future work will focus on the statistical study of untwisting solar jets and their relationship with the kink instability. 

\acknowledgments
We thank the referee for his thorough comments and suggestions. We acknowledge the use of observations from {\it SDO} and {\it IRIS}. Vector magnetic field data is courtesy of the HMI science team. We acknowledge the use of codes in the following papers: \cite{Welsch2003} (YAFTA), \cite{Wiegelmann2008} (NLFFF), and \cite{Schuck2008} (DAVE4VM). JL also benefits from discussions with Dr. Etienne Pariat. JL and RE acknowledge the support received by the Science and Technology Facility Council (STFC), UK. RE is grateful for the support received from the Royal Society (UK). This work is also supported by the Anhui Provincial Natural Science Foundation, China. YW is supported by the grants 41574165 and 41421063 from NSFC.

\bibliographystyle{yahapj}

\end{document}